\documentclass[aps,prl,amsfonts,amsmath,superscriptaddress,floatfix,twocolumn,nofootinbib]  {revtex4-2}

\usepackage{graphicx}
\usepackage{dcolumn}
\usepackage{bm}
\usepackage{hyperref}
\usepackage[dvipsnames]{xcolor}
\preprint{APS/123-QED}
\begin{document}
\title{A generic mechanism for shell structure evolution from an EDF perspective}

\author{L. Heitz}
\affiliation{IJCLab, Universit\'e Paris-Saclay, CNRS/IN2P3, 91405 Orsay Cedex, France}
\affiliation{IRFU, CEA, Universit\'e Paris-Saclay, 91191 Gif-sur-Yvette, France}a
\affiliation{CEA,DAM,DIF, F-91297 Arpajon, France}
\affiliation{Universit\'e Paris-Saclay, CEA, Laboratoire Mati\`ere en Conditions Extr\^emes, 91680, Bruy\`eres-le-Ch\^atel, France}
\author{J.-P. Ebran}
\affiliation{CEA,DAM,DIF, F-91297 Arpajon, France}
\affiliation{Universit\'e Paris-Saclay, CEA, Laboratoire Mati\`ere en Conditions Extr\^emes, 91680, Bruy\`eres-le-Ch\^atel, France}
\author{E. Khan}
\affiliation{IJCLab, Universit\'e Paris-Saclay, CNRS/IN2P3, 91405 Orsay Cedex, France}
\affiliation{Institut Universitaire de France (IUF)}
\author{D.Verney}
\affiliation{IJCLab, Universit\'e Paris-Saclay, CNRS/IN2P3, 91405 Orsay Cedex, France}
\date{\today}

\begin{abstract}
A simple pattern of organisation, the nuclear shell structure, emerges from the complex interactions between nucleons in nuclei and determines, to some significant degree, nuclear structure properties.  
Recent experimental investigations of exotic nuclei revealed a shortfall in our current understanding of nuclear shell evolution and nuclear magicity.
We introduce a novel perspective in the Energy Density Functional framework, where the Dirac mass kinetic term, which stems from the singular participation of a spin-0 boson in the nuclear strong force, plays a pivotal role in generating the nuclear shell structure. Namely, the combination of the Dirac mass kinetic term with the spin-orbit term redefines magic numbers both in stable and exotic nuclei. The identification of this mechanism allows to provide a broad understanding of the origin and evolution of nuclear magic numbers. 
\end{abstract}

\maketitle


\section{Introduction}

The appearance of magic numbers in nuclei, and their evolution far from stability, remains a central problem in nuclear structure physics. In the traditional independent-particle picture, magic numbers are related to large energy gaps in the single-particle spectrum, a pattern commonly attributed to the combined action of a central mean-field potential and a strong spin–orbit interaction \cite{mayer_closed_1949,bm}.

The evolution of these shell closures away from stability has been investigated using a variety of theoretical approaches. Historically, empirical shell-model studies have played a central role, where modifications of effective interactions—most notably tensor components—have been invoked to account for the appearance or erosion of magic numbers in exotic nuclei \cite{otsuka_magic_2001,otsuka_evolution_2005,caurier2005,otsuka_evolution_2020,nowacki_neutron-rich_2021}. From a phenomenological point of view, the possible role of the evolution pseudo-spin partners was noticed in the conclusions of \cite{sorlin_nuclear_2008}, calling for a deeper understanding of this effect. More recently, the emergence of magicity has also been addressed from an \textit{ab initio} perspective, where shell evolution provides a stringent test of chiral interactions and many-body frameworks \cite{miyagi2020,heinz2025,ding25}.

Energy Density Functional (EDF) theory provides an alternative and global framework in which these questions can be revisited. There are EDFs studies of magicity \cite{rodriguezguzman2002,rodriguez2003,ohta2005,bender2008, rodriguez2016}, but there is no systematic overview on this topic. In EDF approaches, two distinct levels of description coexist \cite{bender_self-consistent_2003}. At the single-reference (SR) level, the ground state is approximated by a single Hartree-Fock-Bogliubov (HFB) state, that encodes the single-particle sequence generated by the functional. Although SR-EDF does not yield quantitative separation energies or excitation spectra, it provides the theoretical basis that underlies shell evolution in the EDF framework: features of the SR single-particle structure determine the quality and behavior of subsequent multi-reference (MR) calculations. In MR-EDF, collective correlations are incorporated by mixing different SR quasiparticle vacua, but the predictive power of the method ultimately correlates on the robustness of the underlying SR single-particle patterns.

This motivates a systematic analysis of the single-particle structure generated by EDFs, with a particular focus on the mechanisms that could unify our understanding of the evolution of magic numbers across the nuclear chart. Therefore, the central questions we address, within an EDF formulation, are: is there a common microscopic mechanism driving the emergence and migration of shell closures? Which interaction terms are required to capture this mechanism in a reliable and transferable manner?

Our study adopts this perspective and concentrates on the qualitative trends encoded at the SR-EDF level, which act as precursors to the quantitative results obtained in MR-EDF treatments. By contrasting relativistic and non-relativistic functionals, we specifically investigate how their respective single-particle structures—particularly the behavior of pseudo-spin partners—shape the appearance of magic numbers and their evolution along isotopic and isotonic chains. We illustrate both of the points by examining the formation of the neutron magic number 28 in a stable isotope; then by studying the evolution of orbitals that participate in the creation of the neutron magic number 50. We propose a unified and general mechanism according to which the combined effect of the spin-orbit term and the hereafter so-called Dirac mass kinetic term shows good qualitative consistency with empirical evidences, and applies to cases of current experimental focus. It should be noted that, single-particle energies, though not observables ~\cite{BARANGER1970225,duguet_2012,duguet_2015}, remain an essential diagnostic for comparing EDF predictions, as they are strongly correlated to the quality of the approach.

\begin{figure}
\includegraphics[width=\linewidth]{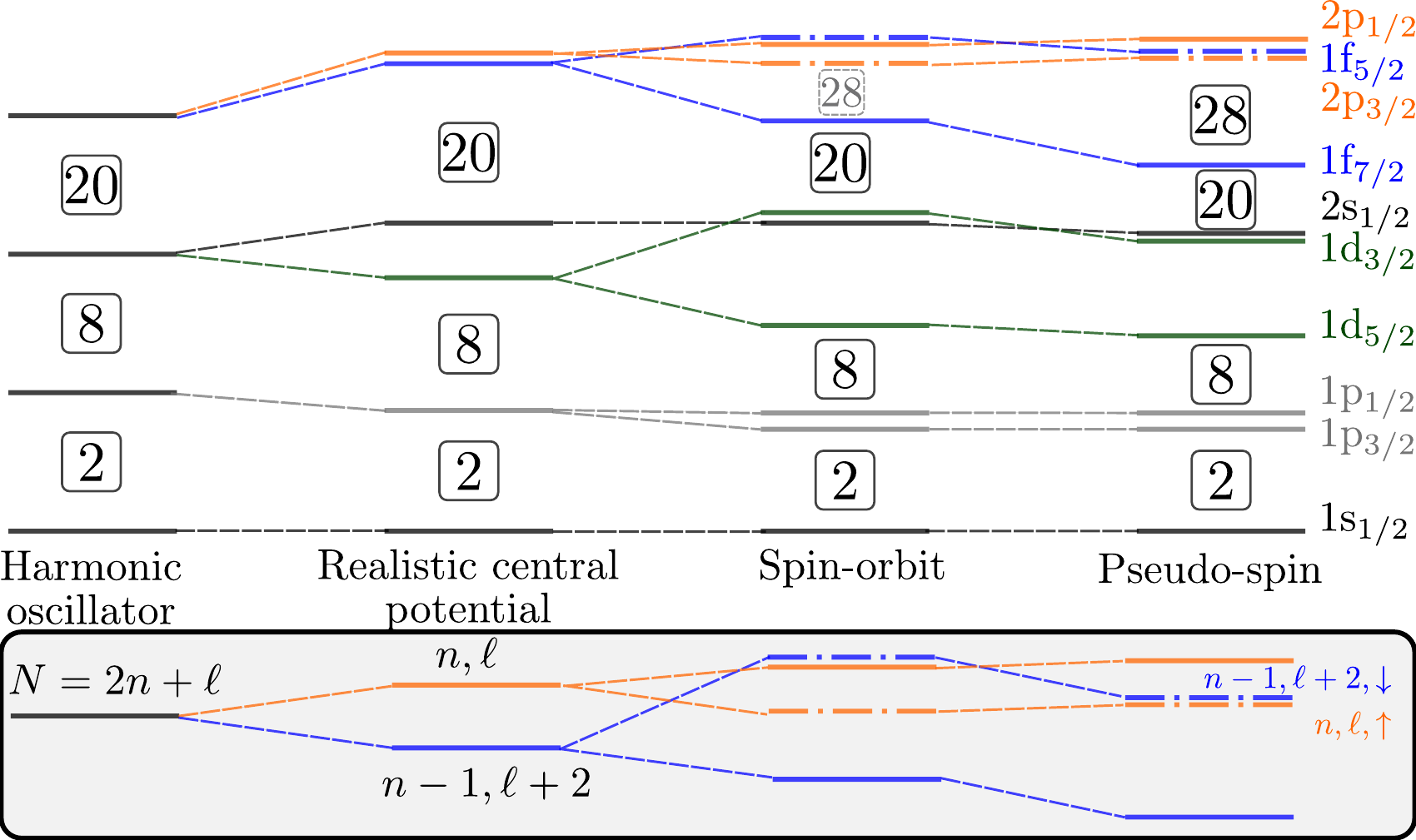}
\caption{\label{fig:fig1} (upper panel) $^{56}$Ni (Z=28, N=28) shell structure induced by the main components of the nuclear confining potential . The single particle energies are calculated within the relativistic mean field perturbative framework described in \cite{liang_perturbative_2011}, the reference state being a relativistic harmonic oscillator. The functional used is DD-MEV. \cite{mercier_covariant_2023}.}
\end{figure}

\section{Covariant EDF framework and the Dirac kinetic term}

A central question in an EDF analysis is to identify all of the dominant contributions to the mean-field potential and how they interfere with each other. In particular, the way spin-related symmetries are realized (or broken) plays an important role in shaping the nuclear shell structure. In that regards, Covariant Energy Density Functionals (cEDFs) provide a natural setting in which these symmetries are embedded, for three reasons: (i) they are based on the Lorentz group, hence spin-related symmetries are naturally built-in without any prior assumptions, (ii) the form of medium-dependent 2-body forces captures effects beyond pairwise interactions and (iii) they can access nuclei presently beyond the reach of  fully converged ab initio calculations. These features make them a useful tool at the SR level for analyzing qualitative trends in shell evolution. In this work, we use spherical Relativistic Hartree-Bogoliubov calculations~\cite{niksic_dirhb_2014,tian_finite_2009}, with the DD-MEV parametrization~\cite{mercier_covariant_2023}, as a qualitative probe for the underlying EDF single-particle structure.
In this framework, the central and spin-dependent components are subsumed in the form of a one-meson-exchange. Standard parameterizations, such as DD-MEV, include the minimal set of meson fields yielding a quantitative description of finite nuclei and infinite nuclear matter properties, i.e a spin-0 meson, $\sigma$, whose exchange generates the attractive part of the nucleon-nucleon interaction, and two spin-1 mesons, $\omega$ and $\rho$, whose exchange produces the short-range repulsive part of the nucleon-nucleon interaction. 

In order to compare these covariant EDF with their non-relativistic counterparts, a non-relativistic limit of the cEDF approach is taken. This yields an equivalent non-relativistic mean-field Hamiltonian $H$, where all the dominant contributions to the mean-field potential are included and where all spin-related symmetries are properly accounted for. Such a non-relativistic reduction can be performed, e.g., via a similarity renormalization group transformation~\cite{huang_investigation_2022,shi_examination_2014,guo_nonrel_2019}, yielding, up to first order in $1/M^2$, with $M$ the nucleon mass:

\begin{equation}
\label{eq:H}
    H = H_0 + H_{\rm{so}} + H_{\rm{DKT}},
\end{equation}
\noindent where 
\begin{eqnarray}
\label{eq:H0} H_0 &=& \frac{p^2}{2M} + (V+S)(r),\\
\label{eq:Hso} H_{\rm{so}} &=& - \frac{\kappa}{r}\frac{V'-S'}{4M^2}, \\
\label{eq:HDKT} H_{\rm{DKT}} &=& -\frac{1}{2M^2}\left[pS(r)p\right].
\end{eqnarray}

and $V'$ and $S'$ stand for the spatial derivative of $V$ and $S$, respectively.

\noindent It should be noted that standard non-relativistic Skyrme and Gogny parameterizations typically retain only the two first terms of Eq.~\eqref{eq:H}, i.e. $H_0$ and $H_{\rm{so}}$. $H_0$ comprises the kinetic ($p$ represents the linear momentum of a nucleon) and central potential terms, where $S$ ($V$) is the attractive (repulsive) mean-potential generated by the exchange of a spin-0 (spin-1) boson, with a typical potential depth $S\sim -400$ MeV ($V\sim 350$ MeV)~\cite{ebran_spinorbit_2016}. 
$H_{\rm{so}}$ stands for the spin-orbit term that naturally emerges with a large magnitude~\cite{ebran_spinorbit_2016}. It features the quantity $\kappa \equiv (\ell-j)(2j+1)$, $\ell$ being the angular momentum and $j$ the total angular momentum. An additional contribution of comparable magnitude also emerges, $H_{\rm{DKT}}$, hereafter called the Dirac mass Kinetic Term (DKT). Although already known in cEDFs \cite{huang_investigation_2022,shi_examination_2014,guo_nonrel_2019}, its effect on nuclear shell structure has not been systematically analyzed yet. The DKT is connected with the presence of a spin-0 mediator via the mean potential $S(r)$. Into non-relativistic language, it translates the renormalization of the nucleon mass due to the presence of a spin-0 field.  
The absence of a counterpart in atomic physics (which only involves a spin-1 mediator - the photon -) may help explain why its global impact on nuclear shell structure has received relatively little attention so far.

\section{Formation of the N=28 shell closure}

Having set the general framework of the EDF method, we now turn to a concrete illustration by analyzing the formation of the neutron magic number $N=28$. When dealing with a theory involving single-particle spectra, two kinds of spin symmetries are at play: the spin and pseudo-spin ones~\cite{arima_pseudo_1969,hecht_generalized_1969,ginocchio_pseudospin_1997}. These symmetries  originate from the structure of the Dirac Hamiltonian and leave characteristic imprint on shell structure. Indeed, above magic number 8, a "magic" Fermi gap involves a set of orbitals linked by both spin and pseudo-spin symmetries. Spin symmetry connects so-called spin-orbit partners, i.e. orbitals with quantum numbers $(n,\ell, j =\ell+1/2)$ and $(n,\ell, j =\ell-1/2)$, where $n$ is the radial quantum number. In a covariant formulation,  the breaking of spin symmetry is governed by the gradient of $(V-S)(r)$, leading to non-zero spin-orbit splittings, whose magnitudes are governed by the magnitude of $V-S$ \cite{ebran_spinorbit_2016}. Pseudo-spin symmetry connects so-called pseudo-spin-orbit (PSO) partners, i.e. orbitals with quantum numbers $(n,\ell,j=\ell+1/2)$ and $(n-1,\ell+2, j'= j+1)$. In a covariant framework, the commutator of the generator of the pseudo-spin symmetry with the Dirac Hamiltonian is proportional to the gradient of $(S+V)(r)$, the latter being considered as a small quantity ~\cite{arima_pseudo_1969,hecht_generalized_1969,ginocchio_pseudospin_1997}. Depending on the nucleus, and PSO partners at stake, pseudo-spin symmetry may be strongly broken or nearly realized, leading to PSO splitting of varying sign.

This suggests that an explicit DKT-like term may be needed, in a non-relativistic setting, to systematically capture the interplay between spin and pseudo-spin partners, as it appears as a leading contribution in a non-relativistic limit of a covariant EDF. Would that term be omitted, it is likely that both spin symmetries may be captured locally - for instance in the calibration region for the fitting protocol of the functional, by compensation of other terms. However, such a local remedy is likely to fail when extrapolating outside of the fitting region and compensations may no longer hold.

Fig. \ref{fig:fig1} illustrates how the interplay of both the spin and pseudo-spin symmetries affects the formation of the neutron magic number 28 in a stable nucleus, from a covariant EDF calculation.  Without any pseudo-spin effect, the full spin-orbit splitting is already obtained, see for instance the $1f_{7/2}-1f_{5/2}$ splitting in the 3rd column. This is analogous to what would be obtained in a non-relativistic setting - except for the magnitude of the gap. However, in the present covariant framework, the spin-orbit splitting is not enough to generate a pronounced $N=28$ magic gap. Consistently including both spin and pseudo-spin symmetries  (4th column) triggers the formation of the typical large N=28 magic gap: first, the $1f_{5/2}$ and $2p_{3/2}$ orbitals get closer to ensure a correct PSO gap, namely its quasi-degeneracy. This, in turn, shifts the $1f_{7/2}$ level downwards, while maintaining the  
 $1f_{7/2}-1f_{5/2}$ spin-orbit gap.

\section{Evolution of pseudo-spin partners along the N=50 chain}
 
This example suggests that PSO splitting may play a key role in shaping magic numbers, from an EDF formulation. It should also be noted that a recent \textit{ab initio} study also highlights the role of pseudo-spin symmetry in building magic number \cite{ding25}. It should be noted that fitting EDF free parameters to stable nuclei may mask the explicit impact of the DKT, as its effects may partially be incorporated into renormalized central and spin-orbit terms. This helps to explain why the traditional understanding of magic numbers in terms of central and spin-orbit potentials has been successful near stability  \cite{mayer_closed_1949,bm}. However this local compensation does not necessarily hold when extending the analysis to more exotic nuclei: as the PSO splitting is not constant with mass number, a failure of this remedy is likely to happen when extrapolating towards exotic nuclei. In what follows, we examine how the PSO gaps are impacted by the DKT, along an isotonic chain.

\begin{figure}
\includegraphics[width=\linewidth]{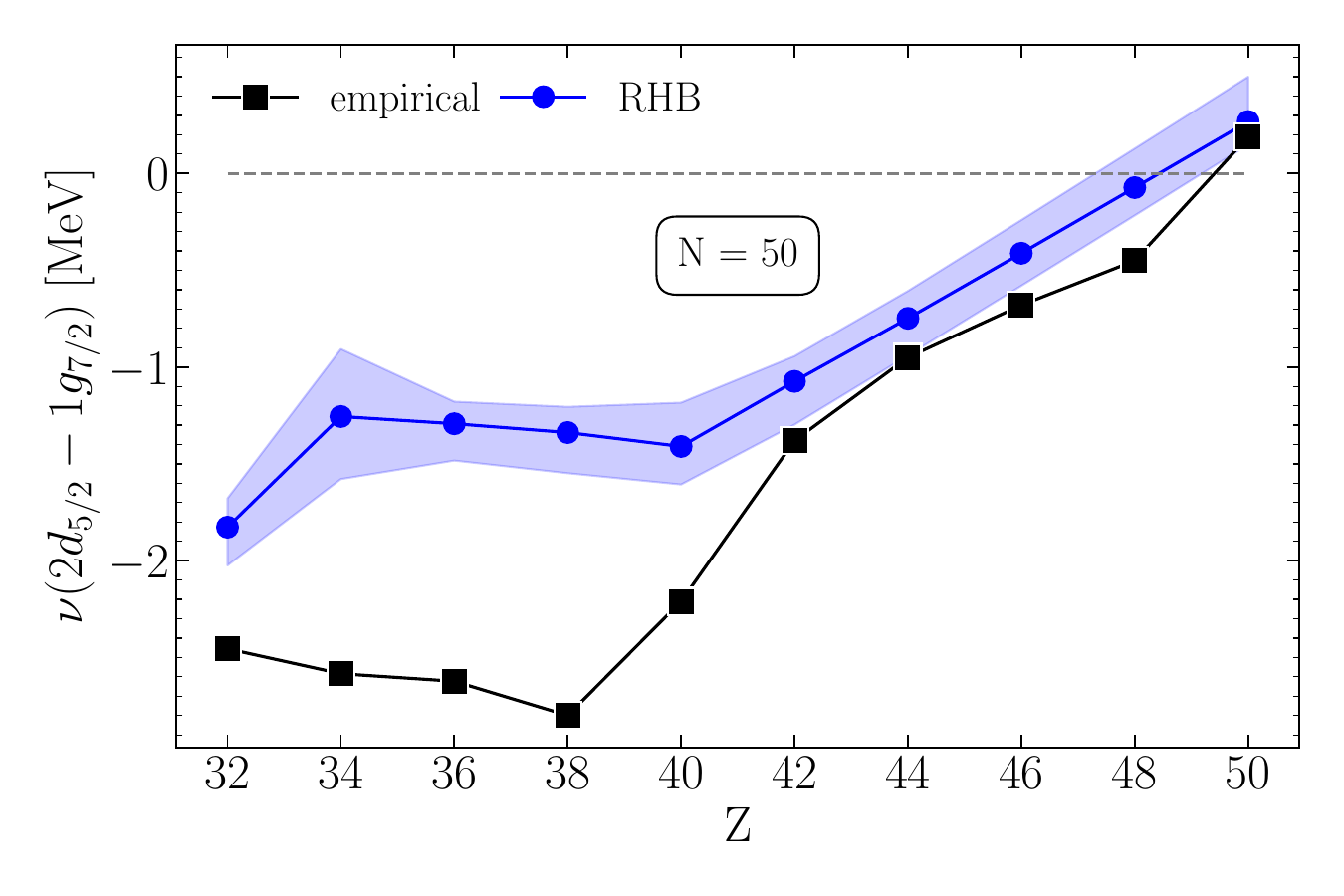}
\caption{\label{fig:fig2}$\nu$(2d$_{5/2}$-1g$_{7/2}$) energy gap in the N=50 chain obtained with several covariant relativistic Hartree-Bogoliubov calculations ("RHB"). The lines within the bands represent  the mean value for the calculation, as averaged over different parametrizations. The empirical data are taken from Ref.~\cite{otsuka_evolution_2020}. See the Supplemental Material \cite{supplemental_material} for details regarding the functionals used.}
\end{figure}

On this purpose, we consider the evolution of magic number $N=50$, which is formed by $1g_{9/2}$ and the lowest of the two PSO partners $(1g_{7/2},2d_{5/2})$.
Fig. \ref{fig:fig2} displays the evolution in the N=50 isotonic chain of the PSO gap between the $2d_{5/2}$ and $1g_{7/2}$ neutron orbitals, involved in the formation of the neutron magic number 50. *
An important benchmark for modeling the $N=50$ chain is the experimentally observed level inversion between the first $5/2^+$ and $7/2^+$ excited states in $^{100}$Sn ($Z=50$) \cite{darby2010}. In a single-particle picture, this should be traced back to an inversion between $\nu 2d_{5/2}$ and $\nu 1g_{7/2}$ states, which are PSO partners. In other words, the PSO gap should change its sign around $Z=50$. Different RMF parametrizations have been tested (see \cite{supplemental_material}), and all exhibit a crossing, as shown in Fig.~\ref{fig:fig2}, although the corresponding $Z$ value depends on the parametrization used. It should be noted that most RMF parametrizations considered here do not include a tensor force, and nevertheless reproduce the crossing.
The role of the DKT in achieving this crossing is further emphasized in the Supplemental Material \cite{supplemental_material}, where the PSO gap is decomposed into the contribution coming from the DKT alone and that from the remaining components, showing that its inclusion is decisive to achieve the crossing between PSO partners. Similar trends are observed for other PSO gaps, e.g. the ones involved in the formation of the proton magic numbers 28 and 50~\cite{delafosse_pseudospin_2018}.  
Still, the importance of the tensor force for describing PSO gaps has been highlighted in several studies \cite{li2016a,karakatsanis_spin-orbit_2017,long_relativistic_2022}. However, its effect appears to be more pronounced in heavier systems than those considered here. For instance, the $h_{11/2}$–$g_{9/2}$ repulsion in tin isotopes cannot be reproduced with the present models, suggesting that tensor contributions may play a smaller role for the nuclei studied in the present work.
It should also be noted that most NR interactions fail to reproduce this crossing (see \cite{supplemental_material}). Some specific Skyrme interactions, such as SIII complemented with a tensor term \cite{brink2018}, are nevertheless able to reproduce the inversion of the PSO partners around $Z=50$.

\begin{figure*}
\includegraphics[width=\linewidth]{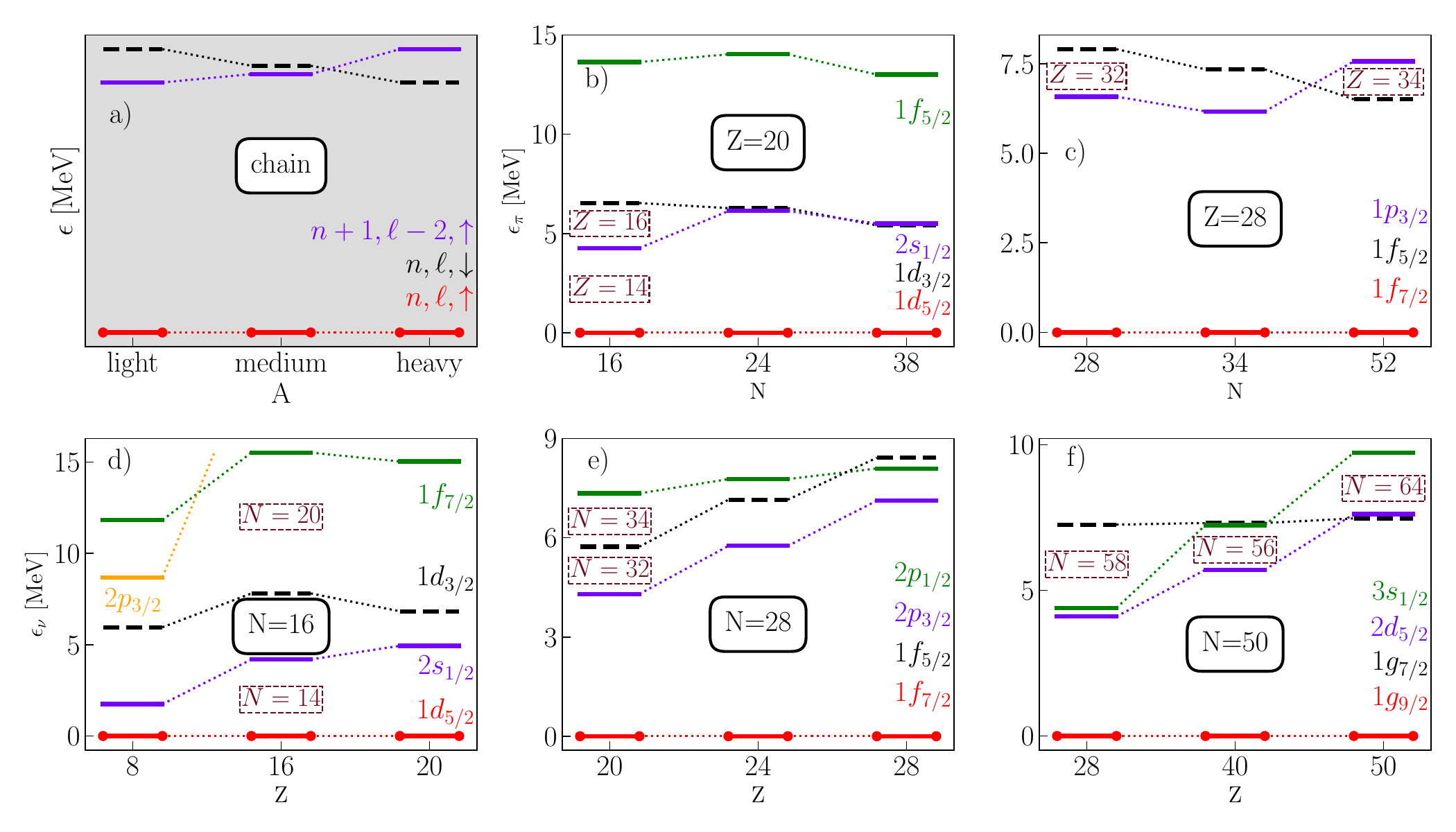}
\caption{\label{fig:fig3}Evolution of shell structure in isotopic and isotonic chains, using relativistic Hartree-Bogoliubov calculations with DD-MEV functional \cite{mercier_covariant_2023} (see text). As the mass increases, the orange and blue orbitals eventually become degenerate, or even cross. These orbitals are the two PSO partners, and the corresponding SSE mechanism accounts for the appearance of new magic numbers. }
\end{figure*}

\section{Generalization : a generic Spin Symmetries Effect mechanism}

A unified and generic mechanism is emerging from both empirical evidences and the present SR-cEDF calculation. It is hereafter called the Spin Symmetries Effect (SSE). It shows how the evolution of PSO gaps shapes nuclear magicity, and is sketched in panel a) of Fig. \ref{fig:fig3}. Three orbitals, connected by two types of relations, are involved in the SSE. The orbital with quantum numbers $(n,\ell, j=\ell-1/2)$, is at the same time the spin-orbit partner of the $(n,\ell, j+1)$ orbital and the PSO partner of the $(n+1,\ell-2, j-1)$ orbital. For instance, $(n=1, \ell =3, j = 5/2)\equiv 1f_{5/2}$ is the spin-orbit partner of $1f_{7/2}$ and the pseudo-spin orbit partner of $(n+1=2,\ell-2=1, j'=1/2) \equiv 2p_{3/2}$. In the lightest nuclei of a given chain, the PSO partners are well separated in energy, the $j$ orbital lying above the $j-1$ one. As the number of nucleons increases, the spin-orbit gap only slightly changes, while the PSO gap grows from a negative to a positive value. The present work provides a microscopic explanation for the phenomenological PSO effect noticed in \cite{sorlin_nuclear_2008}: it naturally emerges from the cEDF calculation, without any additional force. It should be noted that such an evolution of PSO gaps in Ca, Zr, and Sn isotopes has been previously studied \cite{Meng_1999,LOPEZ_2018}, showing a global decrease of its absolute value and its inversion. However, the combined evolution of SO and PSO splittings, which together generate the SSE patterns, has not been discussed in a unified framework in previous works. How the SSE operates can now be illustrated in more detail in cases of current experimental focus. Two remarks are in order. First, it should be noted that pairing correlations are taken into account, as the calculations are performed at the relativistic Hartree–Bogoliubov (RHB) level. Second, the mechanism at play, which is described within the RMF framework, is more general as it relies solely on the empirical determination of single-particle levels.

The evolution of magicity in N=16 isotones is displayed in panel d) of figure \ref{fig:fig3}. The isotonic chain is not long enough for the $1d_{3/2}$ and $2s_{1/2}$ PSO partners to cross. This is consistent with a persistence of the $N=16$ shell closure from $^{24}$O to $^{36}$Ca, and in agreement with the recent measurements of Ref. \cite{lalanne_n_2023}.  On the other hand, the SSE leads to a large $2p_{3/2}-1f_{5/2}$ (not shown) PSO gap dropping $2p_{3/2}$ below $1f_{7/2}$ around the Oxygen isotopes, and reducing the magic gap N=20. The calculation along the $N = 20$ isotonic chain can be found in Supplemental Material \cite{supplemental_material}. In the case of the Ca isotopic chain (panel b), Z = 20 is found to be a robust magic number, again in agreement with experimental data  \cite{kanungo_observation_2002}.

The situation is slightly different for Nickel isotopes (panel c)). The crossing of the $2p_{3/2}$ and $1f_{5/2}$ PSO partners is related to the appearance of a $Z=32$ proton subshell closure below $^{62}$Ni. This crossing also explains the Z=32 erosion beyond N = 50, favoring the appearance of the new proton subshell closure 34, as flagged by the experimental data \cite{flanagan_nuclear_2009}. This illustrates the decisive effect of level crossing, as discussed through Fig. \ref{fig:fig2}. It should be noted that the Z = 28 gap is predicted to be robust over the Nickel chain.

The N = 28 isotones (panel e) provide an example of interplay between the spin and pseudo-spin properties. Around $^{48}$Ca, the ordering of the $2p_{1/2}$ and $2p_{3/2}$ spin-orbit partners and of the $2p_{3/2}$ and $1f_{5/2}$ PSO partners is such that two subshell closures emerge around $N=32 $ and $N=34$, respectively. By virtue of the SSE, as Z increases, the PSO partners get closer while the spin-orbit gap stays relatively constant. This leads to a slight reduction of the N=32 gap and to a significant weakening of the $N=34$ closure. This is again consistent with recent experimental evidence of N=34 being significant only around Z = 20 \cite{iimura_study_2023}. 

Finally, another instance of the SO/PSO interplay occurs in the N=50 isotonic chain (panel f), with $3s_{1/2}$ being the PSO partner of $2d_{3/2}$ (not shown in this figure for simplicity). Its spin-orbit partner, $2d_{5/2}$, is the PSO partner of $1g_{7/2}$. The $1g_{7/2}-2d_{5/2}$ and $2d_{3/2}-3s_{1/2}$ PSO gaps are large in absolute value, such that $3s_{1/2}$ and $2d_{5/2}$ are accidentally quasi degenerate around $^{78}$Ni. This configuration leads, in the calculation, to a reduced $N=50$ magic gap around Z = 28, in agreement with the data, see Ref. \cite{thisse_study_2023} and references therein.  
According to the SSE, as Z increases, the $1g_{7/2}-1g_{9/2}$ spin-orbit gap is almost unchanged whereas the $1g_{7/2}-2d_{5/2}$ and $2d_{3/2}-3s_{1/2}$ PSO gaps are reduced in absolute value:  $1g_{7/2}$ and $2d_{5/2}$ become degenerate around $^{100}$Sn, and $3s_{1/2}$ crosses $1g_{7/2}$ around $^{90}$Zr.
This behavior suggests an erosion of the N = 56 closure \cite{TONDEUR1981278} as Z decreases, when approaching $^{78}$Ni (Z = 28). It points to the possible emergence of an N = 58 subshell gap of comparable size. This result provides a robust prediction for nuclei that remain highly challenging to access experimentally. However, early indirect corresponding indications were noted for Z $\approx$ 31 \cite{Winger_2010} and more recent observations in the region Z $\approx$ 35, 56 $\leq$ N $\leq$ 60 appear to support these findings \cite{dudouet_kr_2017,Dudouet2024}. The disappearance of the N = 56 gap while approaching $Z=50$, in favor of the emergence of the N = 64 subshell closure, is also in agreement with available experimental indications \cite{Kumar2010}. These claims have been verified by computing neutron single particle energies for $^{86}$Ni and $^{114}$Sn, see Supplemental Material \cite{supplemental_material}).

\begin{figure*}
\includegraphics[width=\linewidth]{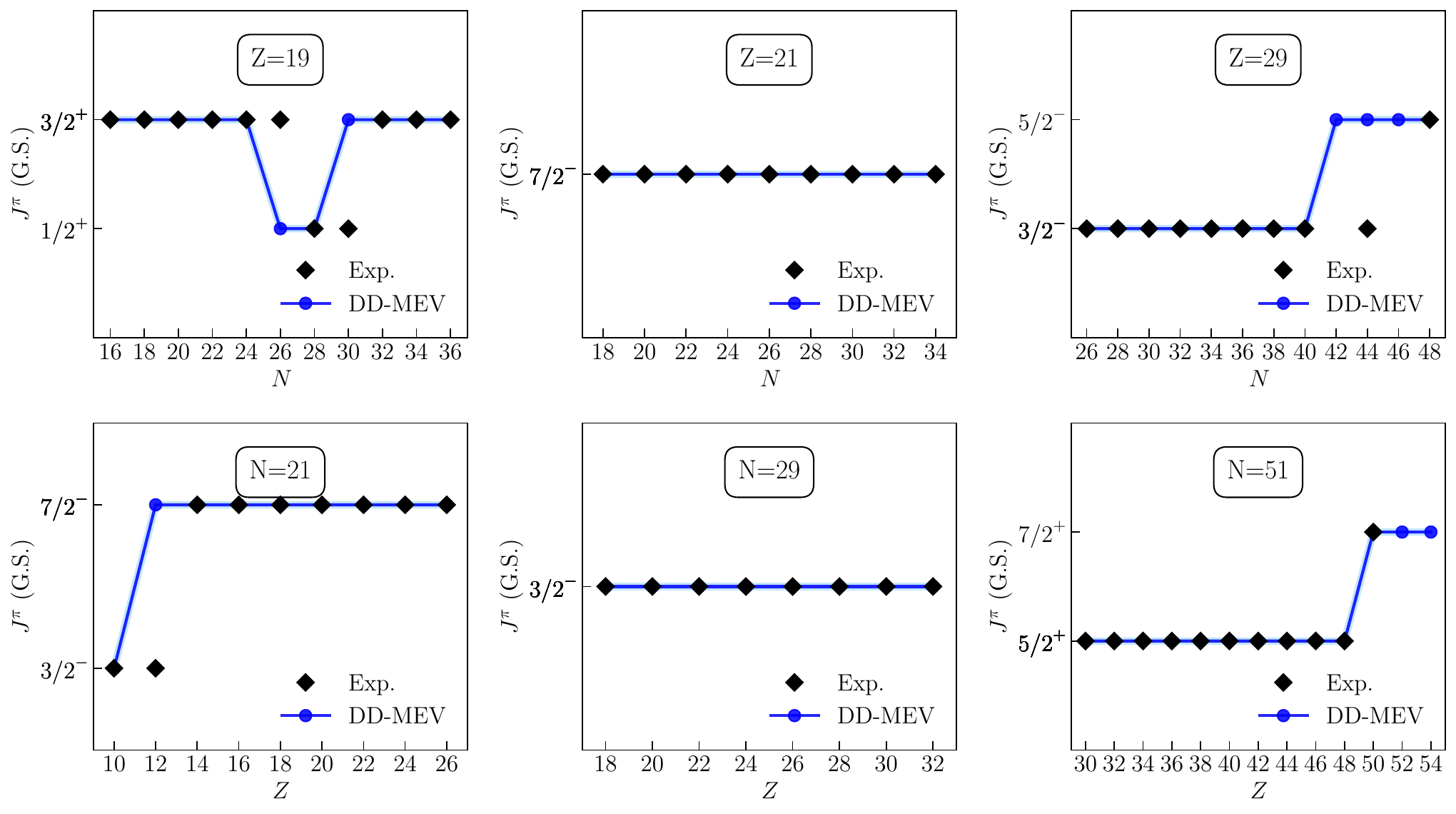}
\caption{\label{fig:fig4}Evolution of odd-even (or even-odd) ground state spin-parity, as predicted by mere filling of even-even RHB calculation with the DD-MEV parametrization,  as compared to  experimental data. Experimental data is taken from \cite{ensdf2025}.}
\end{figure*}

The previous claims are supported by comparing the expected spin–parities of neighboring odd–even (even–odd) nuclei with those predicted from a mere filling of the lowest unoccupied state in even–even RHB calculations with the DD-MEV parametrization, as shown in Fig.~\ref{fig:fig4}. A PSS level inversion is expected to correlate with a spin–parity inversion in the neighboring nuclei.
In particular, for $N=50$, the $\nu(2d_{5/2}-1g_{7/2})$ inversion is reflected in the change of $J^\pi$ from $5/2^+$ to $7/2^+$ in the $N=51$ isotones. The same behaviour is observed for the $Z=28$ and $N=20$ chains. 
On the other hand, the near degeneracy, and even crossing, of $\pi(2s_{1/2}-1d_{3/2})$ in the $Z=20$ chain leads to a change in spin–parity in $Z=19$ nuclei around $N=28$, which is corroborated by experimental data. 
Conversely, the absence of a level crossing above the magic gap in the $Z=20$ and $N=28$ chains results in a constant $J^\pi$ along the $Z=21$ and $N=29$ chains, respectively.

\section{Conclusion}

In summary, the present work highlights a recurring pattern, the Spin Symmetries Effect, that appears to underlie trends in both the emergence and evolution of magicity in nuclei, based on the interplay between spin-orbit and pseudo spin-orbit partners. A qualitatively sound description of these trends can be obtained when the mean-field potential includes explicit central, spin-orbit and Dirac Kinetic terms, the latter arising from the specific scalar (spin-0) channel of the relativistic mean-field. Obtaining a fully quantitative description of shell evolution further requires a multi-reference realization of the EDF, for more correlations to be captured; and when reaching heavier masses, the inclusion of tensor force may also becomes significant.

 \paragraph{Acknowledgements} The authors acknowledge the participants of the ESNT workshop “Effective yet elusive: exploring single-particles”, for helping clarifying the use of single-particle concepts

\bibliography{apssamp}

@article{bender2008,
  title = {Collectivity-induced quenching of signatures for shell closures},
  author = {Bender, M. and Bertsch, G. F. and Heenen, P.-H.},
  journal = {Phys. Rev. C},
  volume = {78},
  issue = {5},
  pages = {054312},
  numpages = {7},
  year = {2008},
  month = {Nov},
  publisher = {American Physical Society},
  doi = {10.1103/PhysRevC.78.054312},
  url = {https://link.aps.org/doi/10.1103/PhysRevC.78.054312}
}

@article{BARANGER1970225,
title = {A definition of the single-nucleon potential},
journal = {Nuclear Physics A},
volume = {149},
number = {2},
pages = {225-240},
year = {1970},
issn = {0375-9474},
doi = {https://doi.org/10.1016/0375-9474(70)90692-5},
url = {https://www.sciencedirect.com/science/article/pii/0375947470906925},
author = {Michel Baranger}
}

@article{ding25,
  title = {From spin to pseudospin symmetry: The origin of magic numbers in nuclear structure},
  author = {Ding, C. R. and Wang, C. C. and Yao, J. M. and Hergert, H. and Liang, H. Z. and Bogner, S. K.},
  journal = {Phys. Rev. Lett.},
  pages = {--},
  year = {2025},
  month = {Dec},
  publisher = {American Physical Society},
  doi = {10.1103/8lzc-j1lx},
  url = {https://link.aps.org/doi/10.1103/8lzc-j1lx}
}

@misc{supplemental_material,
    note = {See Supplemental Material \url{[url]} for DKT contribution, details regarding functionals and $^{86}$Ni,$^{114}$Sn and N=20 isotonic chain calculations},
}

@article{long_relativistic_2022,
	title = {Relativistic {Hartree}–{Fock} model and its recent progress on the description of nuclear structure $^{\textrm{*}}$},
	volume = {74},
	issn = {0253-6102, 1572-9494},
	url = {https://iopscience.iop.org/article/10.1088/1572-9494/ac70ae},
	doi = {10.1088/1572-9494/ac70ae},
	language = {en},
	number = {9},
	urldate = {2023-04-17},
	journal = {Communications in Theoretical Physics},
	author = {Long, W H and Geng, J and Liu, J and Wang, Z H},
	month = sep,
	year = {2022},
	pages = {097301},
}

@article{duguet_2015,
  title = {Nonobservable nature of the nuclear shell structure: Meaning, illustrations, and consequences},
  author = {Duguet, T. and Hergert, H. and Holt, J. D. and Som\`a, V.},
  journal = {Phys. Rev. C},
  volume = {92},
  issue = {3},
  pages = {034313},
  numpages = {15},
  year = {2015},
  month = {Sep},
  publisher = {American Physical Society},
  doi = {10.1103/PhysRevC.92.034313},
  url = {https://link.aps.org/doi/10.1103/PhysRevC.92.034313}
}

@article{duguet_2012,
  title = {Ab initio approach to effective single-particle energies in doubly closed shell nuclei},
  author = {Duguet, T. and Hagen, G.},
  journal = {Phys. Rev. C},
  volume = {85},
  issue = {3},
  pages = {034330},
  numpages = {13},
  year = {2012},
  month = {Mar},
  publisher = {American Physical Society},
  doi = {10.1103/PhysRevC.85.034330},
  url = {https://link.aps.org/doi/10.1103/PhysRevC.85.034330}
}

@article{li2016a,
  title = {Magicity of Neutron-Rich Nuclei within Relativistic Self-Consistent Approaches},
  author = {Li, Jia Jie and Margueron, J{\'e}r{\^o}me and Long, Wen Hui and Van Giai, Nguyen},
  year = {2016},
  month = feb,
  journal = {Physics Letters B},
  volume = {753},
  pages = {97--102},
  issn = {03702693},
  doi = {10.1016/j.physletb.2015.12.004},
  urldate = {2024-05-07},
  langid = {english},
  language = {en}
}

@article{ginocchio_pseudospin_1997,
	title = {Pseudospin as a {Relativistic} {Symmetry}},
	volume = {78},
	doi = {10.1103/PhysRevLett.78.436},
	language = {en},
	number = {3},
	journal = {Physical Review Letters},
	author = {Ginocchio, Joseph N},
	year = {1997},
	pages = {436--439},
}

@article{tian_finite_2009,
	title = {A finite range pairing force for density functional theory in superfluid nuclei},
	volume = {676},
	issn = {03702693},
	url = {https://linkinghub.elsevier.com/retrieve/pii/S0370269309004912},
	doi = {10.1016/j.physletb.2009.04.067},
	language = {en},
	number = {1-3},
	urldate = {2023-10-24},
	journal = {Physics Letters B},
	author = {Tian, Y. and Ma, Z.Y. and Ring, P.},
	month = jun,
	year = {2009},
	pages = {44--50},
}

@article{niksic_dirhb_2014,
	title = {{DIRHB} -- a relativistic self-consistent mean-field framework for atomic nuclei},
	volume = {185},
	issn = {00104655},
	url = {http://arxiv.org/abs/1403.4039},
	doi = {10.1016/j.cpc.2014.02.027},
	number = {6},
	urldate = {2022-05-02},
	journal = {Computer Physics Communications},
	author = {Niksic, T. and Paar, N. and Vretenar, D. and Ring, P.},
	month = jun,
	year = {2014},
	note = {arXiv: 1403.4039},
	pages = {1808--1821},
}

@article{mercier_covariant_2023,
	title = {Covariant energy density functionals with and without tensor couplings at the {Hartree}-{Bogoliubov} level},
	volume = {107},
	issn = {2469-9985, 2469-9993},
	url = {https://link.aps.org/doi/10.1103/PhysRevC.107.034309},
	doi = {10.1103/PhysRevC.107.034309},
	language = {en},
	number = {3},
	urldate = {2023-10-24},
	journal = {Physical Review C},
	author = {Mercier, F. and Ebran, J.-P. and Khan, E.},
	month = mar,
	year = {2023},
	pages = {034309},
}

@article{thisse_study_2023,
author={Thisse, D.
and Lebois, M.
and Verney, D.
and Wilson, J. N.
and Jovan{\v{c}}evi{\'{c}}, N.
and Rudigier, M.
and Canavan, R.
and Etasse, D.
and Adsley, P.
and Algora, A.
and Babo, M.
and Belvedere, K.
and Benito, J.
and Benzoni, G.
and Blazhev, A.
and Boso, A.
and Bottoni, S.
and Bunce, M.
and Chakma, R.
and Cieplicka-Ory{\'{n}}czak, N.
and Courtin, S.
and Cort{\'e}s, M. L.
and Davies, P.
and Delafosse, C.
and Fallot, M.
and Fornal, B.
and Fraile, L.
and Gjestvang, D.
and Gottardo, A.
and Guadilla, V.
and Gerst, R.-B.
and H{\"a}fner, G.
and Hauschild, K.
and Heine, M.
and Henrich, C.
and Homm, I.
and Hommet, J.
and Ibrahim, F.
and Iskra, {\L} W.
and Ivanov, P.
and Jazrawi, S.
and Korgul, A.
and Koseoglou, P.
and Kr{\"o}ll, T.
and Kurtukian-Nieto, T.
and Meur, L. Le
and Leoni, S.
and Ljungvall, J.
and Lopez-Martens, A.
and Lozeva, R.
and Matea, I.
and Miernik, K.
and Nemer, J.
and Oberstedt, S.
and Paulsen, W.
and Piersa-Silkowska, M.
and Poklepa, W.
and Popovitch, Y.
and Porzio, C.
and Qi, L.
and Ralet, D.
and Regan, P. H.
and Reygadas-Tello, D.
and Rezynkina, K.
and S{\'a}nchez-Tembleque, V.
and Siem, S.
and Schmitt, C.
and S{\"o}derstr{\"o}m, P.-A.
and Solak, K.
and S{\"u}rder, C.
and Tocabens, G.
and Vedia, V.
and Warr, N.
and Wasilewska, B.
and Wiederhold, J.
and Yavahchova, M.
and Zeiser, F.
and Ziliani, S.},
title={Study of {\$}{\$}N=50{\$}{\$}gap evolution around {\$}{\$}Z=32{\$}{\$}: new structure information for {\$}{\$}{\{}{\}}^{\{}82{\}}{\$}{\$}Ge},
journal={The European Physical Journal A},
year={2023},
month={Jul},
day={11},
volume={59},
number={7},
pages={153},
issn={1434-601X},
doi={10.1140/epja/s10050-023-01051-2},
url={https://doi.org/10.1140/epja/s10050-023-01051-2}
}

@article{sorlin_nuclear_2008,
	title = {Nuclear magic numbers: {New} features far from stability},
	volume = {61},
	issn = {01466410},
	shorttitle = {Nuclear magic numbers},
	url = {https://linkinghub.elsevier.com/retrieve/pii/S0146641008000380},
	doi = {10.1016/j.ppnp.2008.05.001},
	language = {en},
	number = {2},
	urldate = {2023-02-17},
	journal = {Progress in Particle and Nuclear Physics},
	author = {Sorlin, O. and Porquet, M.-G.},
	month = oct,
	year = {2008},
	pages = {602--673},
}

@article{shi_examination_2014,
	title = {Examination of the pseudospin symmetry for the relativistic harmonic oscillator with the similarity renormalization group},
	volume = {90},
	issn = {0556-2813, 1089-490X},
	url = {https://link.aps.org/doi/10.1103/PhysRevC.90.034318},
	doi = {10.1103/PhysRevC.90.034318},
	language = {en},
	number = {3},
	urldate = {2023-05-15},
	journal = {Physical Review C},
	author = {Shi, Min and Li, Dong-Peng and Chen, Shou-Wan and Guo, Jian-You},
	month = sep,
	year = {2014},
	pages = {034318},
}

@article{otsuka_evolution_2020,
	title = {Evolution of shell structure in exotic nuclei},
	volume = {92},
	issn = {0034-6861, 1539-0756},
	url = {https://link.aps.org/doi/10.1103/RevModPhys.92.015002},
	doi = {10.1103/RevModPhys.92.015002},
	language = {en},
	number = {1},
	urldate = {2023-04-19},
	journal = {Reviews of Modern Physics},
	author = {Otsuka, Takaharu and Gade, Alexandra and Sorlin, Olivier and Suzuki, Toshio and Utsuno, Yutaka},
	month = mar,
	year = {2020},
	pages = {015002},
}

@article{otsuka_evolution_2005,
	title = {Evolution of {Nuclear} {Shells} due to the {Tensor} {Force}},
	volume = {95},
	issn = {0031-9007, 1079-7114},
	url = {https://link.aps.org/doi/10.1103/PhysRevLett.95.232502},
	doi = {10.1103/PhysRevLett.95.232502},
	language = {en},
	number = {23},
	urldate = {2023-10-03},
	journal = {Physical Review Letters},
	author = {Otsuka, Takaharu and Suzuki, Toshio and Fujimoto, Rintaro and Grawe, Hubert and Akaishi, Yoshinori},
	month = nov,
	year = {2005},
	pages = {232502},
}

@article{otsuka_magic_2001,
	title = {Magic {Numbers} in {Exotic} {Nuclei} and {Spin}-{Isospin} {Properties} of the {NN} {Interaction}},
	volume = {87},
	issn = {0031-9007, 1079-7114},
	url = {https://link.aps.org/doi/10.1103/PhysRevLett.87.082502},
	doi = {10.1103/PhysRevLett.87.082502},
	language = {en},
	number = {8},
	urldate = {2023-10-04},
	journal = {Physical Review Letters},
	author = {Otsuka, Takaharu and Fujimoto, Rintaro and Utsuno, Yutaka and Brown, B. Alex and Honma, Michio and Mizusaki, Takahiro},
	month = aug,
	year = {2001},
	pages = {082502},
}

@article{nowacki_neutron-rich_2021,
	title = {The neutron-rich edge of the nuclear landscape: {Experiment} and theory.},
	volume = {120},
	issn = {01466410},
	shorttitle = {The neutron-rich edge of the nuclear landscape},
	url = {https://linkinghub.elsevier.com/retrieve/pii/S014664102100020X},
	doi = {10.1016/j.ppnp.2021.103866},
	language = {en},
	urldate = {2023-10-04},
	journal = {Progress in Particle and Nuclear Physics},
	author = {Nowacki, Frédéric and Obertelli, Alexandre and Poves, Alfredo},
	month = sep,
	year = {2021},
	pages = {103866},
}

@article{mayer_closed_1949,
	title = {On {Closed} {Shells} in {Nuclei}. {II}},
	volume = {75},
	issn = {0031-899X},
	url = {https://link.aps.org/doi/10.1103/PhysRev.75.1969},
	doi = {10.1103/PhysRev.75.1969},
	language = {en},
	number = {12},
	urldate = {2023-10-04},
	journal = {Physical Review},
	author = {Mayer, Maria Goeppert},
	month = jun,
	year = {1949},
	pages = {1969--1970},
}

@article{lalanne_n_2023,
	title = {N = 16 {Magicity} {Revealed} at the {Proton} {Drip} {Line} through the {Study} of {Ca} 35},
	volume = {131},
	issn = {0031-9007, 1079-7114},
	url = {https://link.aps.org/doi/10.1103/PhysRevLett.131.092501},
	doi = {10.1103/PhysRevLett.131.092501},
	language = {en},
	number = {9},
	urldate = {2023-10-10},
	journal = {Physical Review Letters},
	author = {Lalanne, L. and Sorlin, O. and Poves, A. and Assié, M. and Hammache, F. and Koyama, S. and Suzuki, D. and Flavigny, F. and Girard-Alcindor, V. and Lemasson, A. and Matta, A. and Roger, T. and Beaumel, D. and Blumenfeld, Y and Brown, B. A. and Santos, F. De Oliveira and Delaunay, F. and De Séréville, N. and Franchoo, S. and Gibelin, J. and Guillot, J. and Kamalou, O. and Kitamura, N. and Lapoux, V. and Mauss, B. and Morfouace, P. and Pancin, J. and Saito, T. Y. and Stodel, C. and Thomas, J-C.},
	month = aug,
	year = {2023},
	pages = {092501},
}

@article{kanungo_observation_2002,
	title = {Observation of new neutron and proton magic numbers},
	volume = {528},
	issn = {03702693},
	url = {https://linkinghub.elsevier.com/retrieve/pii/S0370269302012066},
	doi = {10.1016/S0370-2693(02)01206-6},
	language = {en},
	number = {1-2},
	urldate = {2023-11-29},
	journal = {Physics Letters B},
	author = {Kanungo, Rituparna and Tanihata, I and Ozawa, A},
	month = feb,
	year = {2002},
	pages = {58--64},
}

@article{liang_perturbative_2011,
	title = {Perturbative interpretation of relativistic symmetries in nuclei},
	volume = {83},
	issn = {0556-2813, 1089-490X},
	url = {https://link.aps.org/doi/10.1103/PhysRevC.83.041301},
	doi = {10.1103/PhysRevC.83.041301},
	language = {en},
	number = {4},
	urldate = {2022-06-08},
	journal = {Physical Review C},
	author = {Liang, Haozhao and Zhao, Pengwei and Zhang, Ying and Meng, Jie and Giai, Nguyen Van},
	month = apr,
	year = {2011},
	pages = {041301},
}

@article{huang_investigation_2022,
	title = {Investigation of pseudospin and spin symmetries in relativistic mean field theory combined with a similarity renormalization group approach},
	volume = {105},
	issn = {2469-9985, 2469-9993},
	url = {https://link.aps.org/doi/10.1103/PhysRevC.105.054313},
	doi = {10.1103/PhysRevC.105.054313},
	language = {en},
	number = {5},
	urldate = {2023-05-15},
	journal = {Physical Review C},
	author = {Huang, Bo and Guo, Jian-You and Chen, Shou-Wan},
	month = may,
	year = {2022},
	pages = {054313},
}

@article{Winger_2010,
  title = {New subshell closure at $N=58$ emerging in neutron-rich nuclei beyond $^{78}\mathrm{Ni}$},
  author = {Winger, J. A. and Rykaczewski, K. P. and Gross, C. J. and Grzywacz, R. and Batchelder, J. C. and Goodin, C. and Hamilton, J. H. and Ilyushkin, S. V. and Korgul, A. and Kr\'olas, W. and Liddick, S. N. and Mazzocchi, C. and Padgett, S. and Piechaczek, A. and Rajabali, M. M. and Shapira, D. and Zganjar, E. F. and Dobaczewski, J.},
  journal = {Phys. Rev. C},
  volume = {81},
  issue = {4},
  pages = {044303},
  numpages = {7},
  year = {2010},
  month = {Apr},
  publisher = {American Physical Society},
  doi = {10.1103/PhysRevC.81.044303},
  url = {https://link.aps.org/doi/10.1103/PhysRevC.81.044303}
}

@article{LOPEZ_2018,
title = {Tensor force effect on the evolution of single-particle energies in some isotopic chains in the relativistic Hartree-Fock approximation},
journal = {Nuclear Physics A},
volume = {971},
pages = {149-167},
year = {2018},
issn = {0375-9474},
doi = {https://doi.org/10.1016/j.nuclphysa.2018.01.012},
url = {https://www.sciencedirect.com/science/article/pii/S0375947418300198},
author = {M. L\'opez-Quelle and S. Marcos and R. Niembro and L.N. Savushkin}
}

@article{Meng_1999,
  title = {Pseudospin symmetry in Zr and Sn isotopes from the proton drip line to the neutron drip line},
  author = {Meng, J. and Sugawara-Tanabe, K. and Yamaji, S. and Arima, A.},
  journal = {Phys. Rev. C},
  volume = {59},
  issue = {1},
  pages = {154--163},
  numpages = {0},
  year = {1999},
  month = {Jan},
  publisher = {American Physical Society},
  doi = {10.1103/PhysRevC.59.154},
  url = {https://link.aps.org/doi/10.1103/PhysRevC.59.154}
}

@article{guo_nonrel_2019,
  title = {Nonrelativistic expansion of Dirac equation with spherical scalar and vector potentials by similarity renormalization group},
  author = {Guo, Yixin and Liang, Haozhao},
  journal = {Phys. Rev. C},
  volume = {99},
  issue = {5},
  pages = {054324},
  numpages = {9},
  year = {2019},
  month = {May},
  publisher = {American Physical Society},
  doi = {10.1103/PhysRevC.99.054324},
  url = {https://link.aps.org/doi/10.1103/PhysRevC.99.054324}
}

@article{hecht_generalized_1969,
	title = {Generalized seniority for favored {J} ≠ 0 pairs in mixed configurations},
	volume = {137},
	issn = {03759474},
	url = {https://linkinghub.elsevier.com/retrieve/pii/0375947469900773},
	doi = {10.1016/0375-9474(69)90077-3},
	language = {en},
	number = {1},
	urldate = {2022-06-13},
	journal = {Nuclear Physics A},
	author = {Hecht, K.T. and Adler, A.},
	month = nov,
	year = {1969},
	pages = {129--143},
}

@article{iimura_study_2023,
	title = {Study of the {N} = 32 and {N} = 34 {Shell} {Gap} for {Ti} and {V} by the {First} {High}-{Precision} {Multireflection} {Time}-of-{Flight} {Mass} {Measurements} at {BigRIPS}-{SLOWRI}},
	volume = {130},
	issn = {0031-9007, 1079-7114},
	url = {https://link.aps.org/doi/10.1103/PhysRevLett.130.012501},
	doi = {10.1103/PhysRevLett.130.012501},
	language = {en},
	number = {1},
	urldate = {2023-11-15},
	journal = {Physical Review Letters},
	author = {Iimura, S. and Rosenbusch, M. and Takamine, A. and Tsunoda, Y. and Wada, M. and Chen, S. and Hou, D. S. and Xian, W. and Ishiyama, H. and Yan, S. and Schury, P. and Crawford, H. and Doornenbal, P. and Hirayama, Y. and Ito, Y. and Kimura, S. and Koiwai, T. and Kojima, T. M. and Koura, H. and Lee, J. and Liu, J. and Michimasa, S. and Miyatake, H. and Moon, J. Y. and Naimi, S. and Nishimura, S. and Niwase, T. and Odahara, A. and Otsuka, T. and Paschalis, S. and Petri, M. and Shimizu, N. and Sonoda, T. and Suzuki, D. and Watanabe, Y. X. and Wimmer, K. and Wollnik, H.},
	month = jan,
	year = {2023},
	pages = {012501},
}

@article{flanagan_nuclear_2009,
	title = {Nuclear {Spins} and {Magnetic} {Moments} of {Cu} 71 , 73 , 75 : {Inversion} of π 2 p 3 / 2 and π 1 f 5 / 2 {Levels} in {Cu} 75},
	volume = {103},
	issn = {0031-9007, 1079-7114},
	shorttitle = {Nuclear {Spins} and {Magnetic} {Moments} of {Cu} 71 , 73 , 75},
	url = {https://link.aps.org/doi/10.1103/PhysRevLett.103.142501},
	doi = {10.1103/PhysRevLett.103.142501},
	language = {en},
	number = {14},
	urldate = {2023-11-29},
	journal = {Physical Review Letters},
	author = {Flanagan, K. T. and Vingerhoets, P. and Avgoulea, M. and Billowes, J. and Bissell, M. L. and Blaum, K. and Cheal, B. and De Rydt, M. and Fedosseev, V. N. and Forest, D. H. and Geppert, Ch. and Köster, U. and Kowalska, M. and Krämer, J. and Kratz, K. L. and Krieger, A. and Mané, E. and Marsh, B. A. and Materna, T. and Mathieu, L. and Molkanov, P. L. and Neugart, R. and Neyens, G. and Nörtershäuser, W. and Seliverstov, M. D. and Serot, O. and Schug, M. and Sjoedin, M. A. and Stone, J. R. and Stone, N. J. and Stroke, H. H. and Tungate, G. and Yordanov, D. T. and Volkov, Yu. M.},
	month = oct,
	year = {2009},
	pages = {142501},
}

@article{ebran_spinorbit_2016,
	title = {Spin–orbit coupling rule in bound fermion systems},
	volume = {43},
	issn = {0954-3899, 1361-6471},
	url = {https://iopscience.iop.org/article/10.1088/0954-3899/43/8/085101},
	doi = {10.1088/0954-3899/43/8/085101},
	language = {en},
	number = {8},
	urldate = {2022-05-04},
	journal = {Journal of Physics G: Nuclear and Particle Physics},
	author = {Ebran, J-P and Khan, E and Mutschler, A and Vretenar, D},
	month = aug,
	year = {2016},
	pages = {085101},
}

@article{dudouet_kr_2017,
	title = {Kr 36 96 60 –{Low}- {Z} {Boundary} of the {Island} of {Deformation} at {N} = 60},
	volume = {118},
	issn = {0031-9007, 1079-7114},
	url = {http://link.aps.org/doi/10.1103/PhysRevLett.118.162501},
	doi = {10.1103/PhysRevLett.118.162501},
	language = {en},
	number = {16},
	urldate = {2023-12-04},
	journal = {Physical Review Letters},
	author = {Dudouet, J. and Lemasson, A. and Duchêne, G. and Rejmund, M. and Clément, E. and Michelagnoli, C. and Didierjean, F. and Korichi, A. and Maquart, G. and Stezowski, O. and Lizarazo, C. and Pérez-Vidal, R. M. and Andreoiu, C. and De Angelis, G. and Astier, A. and Delafosse, C. and Deloncle, I. and Dombradi, Z. and De France, G. and Gadea, A. and Gottardo, A. and Jacquot, B. and Jones, P. and Konstantinopoulos, T. and Kuti, I. and Le Blanc, F. and Lenzi, S. M. and Li, G. and Lozeva, R. and Million, B. and Napoli, D. R. and Navin, A. and Petrache, C. M. and Pietralla, N. and Ralet, D. and Ramdhane, M. and Redon, N. and Schmitt, C. and Sohler, D. and Verney, D. and Barrientos, D. and Birkenbach, B. and Burrows, I. and Charles, L. and Collado, J. and Cullen, D. M. and Désesquelles, P. and Domingo Pardo, C. and González, V. and Harkness-Brennan, L. and Hess, H. and Judson, D. S. and Karolak, M. and Korten, W. and Labiche, M. and Ljungvall, J. and Menegazzo, R. and Mengoni, D. and Pullia, A. and Recchia, F. and Reiter, P. and Salsac, M. D. and Sanchis, E. and Theisen, Ch. and Valiente-Dobón, J. J. and Zielińska, M.},
	month = apr,
	year = {2017},
	pages = {162501},
}

@article{delafosse_pseudospin_2018,
	title = {Pseudospin {Symmetry} and {Microscopic} {Origin} of {Shape} {Coexistence} in the {Ni} 78 {Region}: {A} {Hint} from {Lifetime} {Measurements}},
	volume = {121},
	issn = {0031-9007, 1079-7114},
	shorttitle = {Pseudospin {Symmetry} and {Microscopic} {Origin} of {Shape} {Coexistence} in the {Ni} 78 {Region}},
	url = {https://link.aps.org/doi/10.1103/PhysRevLett.121.192502},
	doi = {10.1103/PhysRevLett.121.192502},
	language = {en},
	number = {19},
	urldate = {2022-05-09},
	journal = {Physical Review Letters},
	author = {Delafosse, C. and Verney, D. and Marević, P. and Gottardo, A. and Michelagnoli, C. and Lemasson, A. and Goasduff, A. and Ljungvall, J. and Clément, E. and Korichi, A. and De Angelis, G. and Andreoiu, C. and Babo, M. and Boso, A. and Didierjean, F. and Dudouet, J. and Franchoo, S. and Gadea, A. and Georgiev, G. and Ibrahim, F. and Jacquot, B. and Konstantinopoulos, T. and Lenzi, S. M. and Maquart, G. and Matea, I. and Mengoni, D. and Napoli, D. R. and Nikšić, T. and Olivier, L. and Pérez-Vidal, R. M. and Portail, C. and Recchia, F. and Redon, N. and Siciliano, M. and Stefan, I. and Stezowski, O. and Vretenar, D. and Zielinska, M. and Barrientos, D. and Benzoni, G. and Birkenbach, B. and Boston, A. J. and Boston, H. C. and Cederwall, B. and Charles, L. and Ciemala, M. and Collado, J. and Cullen, D. M. and Désesquelles, P. and de France, G. and Domingo-Pardo, C. and Eberth, J. and González, V. and Harkness-Brennan, L. J. and Hess, H. and Judson, D. S. and Jungclaus, A. and Korten, W. and Lefevre, A. and Legruel, F. and Menegazzo, R. and Million, B. and Nyberg, J. and Quintana, B. and Ralet, D. and Reiter, P. and Saillant, F. and Sanchis, E. and Theisen, Ch. and Valiente Dobon, J. J.},
	month = nov,
	year = {2018},
	pages = {192502},
}

@book{bm,
	edition = {W. A. Benjamin},
	title = {Nuclear Structure, Volume 1},
	isbn = {0-8053-1016-9},
	author = {Bohr A, Mottelson B R},
	month = jan,
	year = {1969},
}

@article{bender_self-consistent_2003,
	title = {Self-consistent mean-field models for nuclear structure},
	volume = {75},
	issn = {0034-6861, 1539-0756},
	url = {https://link.aps.org/doi/10.1103/RevModPhys.75.121},
	doi = {10.1103/RevModPhys.75.121},
	language = {en},
	number = {1},
	urldate = {2023-10-19},
	journal = {Reviews of Modern Physics},
	author = {Bender, Michael and Heenen, Paul-Henri and Reinhard, Paul-Gerhard},
	month = jan,
	year = {2003},
	pages = {121--180},
}

@article{arima_pseudo_1969,
	title = {Pseudo {LS} coupling and pseudo {SU3} coupling schemes},
	volume = {30},
	issn = {03702693},
	url = {https://linkinghub.elsevier.com/retrieve/pii/0370269369904432},
	doi = {10.1016/0370-2693(69)90443-2},
	language = {en},
	number = {8},
	urldate = {2023-02-14},
	journal = {Physics Letters B},
	author = {Arima, A. and Harvey, M. and Shimizu, K.},
	month = dec,
	year = {1969},
	pages = {517--522},
}

@article{karakatsanis_spin-orbit_2017,
	title = {Spin-orbit splittings of neutron states in {N} = 20 isotones from covariant density functionals and their extensions},
	volume = {95},
	copyright = {http://link.aps.org/licenses/aps-default-license},
	issn = {2469-9985, 2469-9993},
	url = {https://link.aps.org/doi/10.1103/PhysRevC.95.034318},
	doi = {10.1103/PhysRevC.95.034318},
	language = {en},
	number = {3},
	urldate = {2024-09-12},
	journal = {Physical Review C},
	author = {Karakatsanis, Konstantinos and Lalazissis, G. A. and Ring, Peter and Litvinova, Elena},
	month = mar,
	year = {2017},
	pages = {034318},
}

@article{TONDEUR1981278,
title = {The subshell closure at N = 56 and the deformation of neutron-rich isotopes from Ge to Zr},
journal = {Nuclear Physics A},
volume = {359},
number = {2},
pages = {278-288},
year = {1981},
issn = {0375-9474},
doi = {https://doi.org/10.1016/0375-9474(81)90237-2},
url = {https://www.sciencedirect.com/science/article/pii/0375947481902372},
author = {F. Tondeur}
}

@article{Dudouet2024,
  title = {High-resolution spectroscopy of neutron-rich Br isotopes and signatures for a prolate-to-oblate shape transition at $N=56$},
  author = {Dudouet, J. and Colombi, G. and Reygadas Tello, D. and Michelagnoli, C. and Dao, D. D. and Nowacki, F. and Abushawish, M. and Cl\'ement, E. and Costache, C. and Duch\^ene, G. and Kandzia, F. and Lemasson, A. and Marginean, N. and Marginean, R. and Mihai, C. and Pascu, S. and Rejmund, M. and Rezynkina, K. and Stezowski, O. and Turturica, A. and Ujeniuc, S. and Astier, A. and de Angelis, G. and de France, G. and Delafosse, C. and Deloncle, I. and Gadea, A. and Gottardo, A. and Jones, P. and Konstantinopoulos, T. and Kuti, I. and Le Blanc, F. and Lenzi, S. M. and Lozeva, R. and Million, B. and P\'erez-Vidal, R. M. and Petrache, C. M. and Ralet, D. and Redon, N. and Schmitt, C. and Sohler, D. and Verney, D.},
  journal = {Phys. Rev. C},
  volume = {110},
  issue = {3},
  pages = {034304},
  numpages = {16},
  year = {2024},
  month = {Sep},
  publisher = {American Physical Society},
  doi = {10.1103/PhysRevC.110.034304},
  url = {https://link.aps.org/doi/10.1103/PhysRevC.110.034304}
}

@article{Kumar2010,
  title = {Enhanced ${0}_{\mathrm{g}.\mathrm{s}.}^{+}\ensuremath{\rightarrow}{2}_{1}^{+}$ $E2$ transition strength in $^{112}\mathrm{Sn}$},
  author = {Kumar, R. and Doornenbal, P. and Jhingan, A. and Bhowmik, R. K. and Muralithar, S. and Appannababu, S. and Garg, R. and Gerl, J. and G\'orska, M. and Kaur, J. and Kojouharov, I. and Mandal, S. and Mukherjee, S. and Siwal, D. and Sharma, A. and Singh, Pushpendra P. and Singh, R. P. and Wollersheim, H. -J.},
  journal = {Phys. Rev. C},
  volume = {81},
  issue = {2},
  pages = {024306},
  numpages = {4},
  year = {2010},
  month = {Feb},
  publisher = {American Physical Society},
  doi = {10.1103/PhysRevC.81.024306},
  url = {https://link.aps.org/doi/10.1103/PhysRevC.81.024306}
}

@article{caurier2005,
  author  = {Caurier, E. and Martínez-Pinedo, G. and Nowacki, F. and Poves, A. and Zuker, A. P.},
  title   = {The shell model as a unified view of nuclear structure},
  journal = {Reviews of Modern Physics},
  volume  = {77},
  pages   = {427--488},
  year    = {2005},
  month   = {06},
  doi     = {10.1103/RevModPhys.77.427}
}

@article{miyagi2020,
  author  = {T. Miyagi and S. R. Stroberg and J. D. Holt and N. Shimizu},
  title   = {Ab Initio Multishell Valence-Space Hamiltonians and the Island of Inversion},
  journal = {Physical Review C},
  volume  = {102},
  number  = {3},
  pages   = {034320},
  year    = {2020},
  doi     = {10.1103/PhysRevC.102.034320}
}

@article{heinz2025,
  author  = {Matthias Heinz and Takayuki Miyagi and S. R. Stroberg and A. Tichai and K. Hebeler and A. Schwenk},
  title   = {Improved Structure of Calcium Isotopes from Ab Initio Calculations},
  journal = {Physical Review C},
  volume  = {111},
  number  = {3},
  pages   = {034311},
  year    = {2025},
  doi     = {10.1103/PhysRevC.111.034311}
}

@article{rodriguezguzman2002,
  author  = {R. Rodríguez-Guzmán and J. L. Egido and L. M. Robledo},
  title   = {Correlations Beyond the Mean Field in Magnesium Isotopes: Angular Momentum Projection and Configuration Mixing},
  journal = {Nuclear Physics A},
  volume  = {709},
  number  = {1-4},
  pages   = {201--235},
  year    = {2002},
  doi     = {10.1016/S0375-9474(02)01005-0}
}

\end{document}